\documentclass[aps,prb,reprint]{revtex4-1}
\usepackage{amsmath}
\usepackage{amsfonts}
\usepackage{amssymb}
\usepackage{graphics}
\usepackage{float}
\usepackage{graphicx}
\usepackage{epstopdf}
\usepackage[compact]{titlesec}
\usepackage{natbib}
\usepackage{threeparttable}
\usepackage[titletoc,title]{appendix}
\usepackage{lipsum}
\usepackage{varwidth}
\usepackage{tabularx}
\usepackage{color}
\usepackage{footmisc}  
\usepackage{latexsym}  
\usepackage[none]{hyphenat}  
\usepackage{nicefrac}  
\usepackage{dcolumn}   
\usepackage{bm}        

\newcommand{\ev}{\mathbf{\hat{e}}}

\newcommand{\pv}{\mathbf{p}}
\newcommand{\kv}{\mathbf{k}}
\newcommand{\Rv}{\mathbf{R}}
\newcommand{\Ev}{\mathbf{E}}
\newcommand{\Hv}{\mathbf{H}}

\newcommand{\Gb}{\overline{\overline{\mathbf{G}}}}
\newcommand{\gb}{\overline{\overline{\mathbf{g}}}}
\newcommand{\kb}{\overline{\overline{\mathbf{k}}}}

\newcommand{\epsb}{\overline{\overline{\varepsilon}}}
\newcommand{\mub}{\overline{\overline{\mu}}}
\newcommand{\xib}{\overline{\overline{\xi}}}
\newcommand{\zetab}{\overline{\overline{\zeta}}}

\begin{document}
\title{Electron \textit{g}-factor engineering for non-reciprocal spin photonics}
\author{Parijat Sengupta$^{1}$, Chinmay Khandekar$^{1}$, Todd Van Mechelen$^{1}$, Rajib Rahman$^{2}$ and Zubin Jacob$^{1}$}
\affiliation{$^{1}$Department of Electrical and Computer Engineering \\
Birck Nanotechnology Center \\
Purdue University, West Lafayette, IN 47907, USA \\ $^{2}$School of Physics, The University of New South Wales, Sydney, NSW 2052, Australia}

\begin{abstract}
We study the interplay of electron and photon spin in non-reciprocal materials. Traditionally, the primary mechanism to design non-reciprocal photonic devices has been magnetic fields in conjunction with magnetic oxides, such as iron garnets. In this work, we present an alternative paradigm that allows tunability and reconfigurability of the non-reciprocity through spintronic approaches. The proposed design uses the high-spin-orbit coupling (\textit{soc}) of a narrow-band gap semiconductor (InSb) with ferromagnetic dopants. A combination of the  intrinsic \textit{soc} and a gate-applied electric field gives rise to a strong external Rashba spin-orbit coupling (RSOC) in a magnetically doped InSb film. The RSOC which is gate alterable is shown to adjust the magnetic permeability tensor via the electron \textit{g}-factor of the medium. We use electronic band structure calculations (k$\cdot$p theory) to show the gate-adjustable RSOC manifest itself in the non-reciprocal coefficient of photon fields via shifts in the Kerr and Faraday rotations. In addition, we show that photon spin properties of dipolar emitters placed in the vicinity of a non-reciprocal electromagnetic environment is distinct from reciprocal counterparts. The Purcell factor (F$_{p}$) of a spin-polarized emitter (right-handed circular dipole) is significantly enhanced due to a larger g-factor while a left-handed dipole remains essentially unaffected. Our work can lead to electron spin controlled reconfigurable non-reciprocal photonic devices.
\end{abstract}
\maketitle

\vspace{0.25cm}
\section{Introduction}
\label{s1}
\vspace{0.25cm}
Non-reciprocal photonic materials such as ferrites and magnetized plasmas are central to the design of optical isolators and circulators~\cite{eroglu2010wave}. While technology exists in the microwave regime, there is a major impetus driving on-chip miniaturization of non-reciprocal devices for quantum~\cite{kamal2011noiseless} to classical~\cite{bi2011chip} applications. A particular frontier in this regard is connected to time modulation as a possible pathway to achieve non-reciprocity as an alternative to using magnetic materials. However, significant challenges remain - primarily, insertion loss and the high speed modulation of such effects - which makes it an area of active interest to carry out a search for new materials exhibiting non-reciprocity. 

There is an intimate connection between photon spin~\cite{berry2009optical} and non-reciprocal materials exhibiting gyrotropy. A classical analysis~\cite{kong1975theory} of gyrotropic media reveals that the eigen states of such a medium are circularly polarized with differing phase velocities; however, the role of spin in the near-field of gyrotropic media has not been fully analyzed. In this work, we put forth approaches to probe the near-field spin properties of non-reciprocal media. It is pertinent to note here that the special case of moving media which displays magneto-electric non-reciprocity has fundamental similarities to the Kramers theorem in the near-field regime.~\cite{pendharker2018spin} 

Recently, gyrotropy was demonstrated to be equivalent to effective photon mass  through a direct comparison with an optical-analog of the Dirac equation.~\cite{barnett2014optical,horsley2018topology,van2019photonic}. Gyrotropy, similar to Dirac mass, is accompanied by a low energy (frequency) band gap for propagating waves. Within this band gap, Maxwellian spin waves can exist with unidirectional propagation which are closely related to Jackiw-Rebbi waves that occur at the interface of positive and negative mass media. In addition, the gyro-electric phase of atomic matter combines the principles of non-locality and non-reciprocity to achieve skyrmionic texture of photonic spins in momentum space. This non-local topological electromagnetic phase can host helicity-quantized unidirectional~\cite{van2019unidirectional} edge waves fundamentally different from their classical counterparts - the edge magneto-plasmons. This advancement illustrates how hitherto unexplored forms of gyrotropy can lead to creation of intriguing Maxwellian spin waves as well as spin-1 photonic skyrmionic textures. An equally fundamental application of non-reciprocal materials lies in controlling heat transport~\cite{zhu2018theory,silveirinha2017topological}; thermal energy density in the near-field of a planar slab of gyrotropic media has been predicted to show unidirectional transport behavior even under equilibrium conditions~\cite{khandekar2019thermal}. This effect arises from universal spin-momentum locking of evanescent waves~\cite{van2016universal,bliokh2014extraordinary} in the near-field of a non-reciprocal slab.

The focus of this paper is electron-spin control of gyrotropy which has the potential to utilize spintronic devices with applications requiring photonic non-reciprocity.~\cite{caloz2018electromagnetic} Typically, conventional gyro-electric media rely on cyclotron orbits and orbital angular momentum of electrons interacting with a fixed magnetic field; gyro-magnetic media, on the other hand, obtain their non-reciprocal behavior from electron spin angular momentum interaction with the static magnetic bias. These materials are also widely known as magneto-optic media. Here, we couple band structure calculations - performed using an eight-band k$\cdot$p Hamiltonian adapted~\cite{sengupta2016numerical}  to quantum wells - to the theory of magnetic permeability tensors. This leads to a computation of the non-reciprocity coefficient inside matter for photon fields.   

We propose nanoscale thick InSb quantum well structures~\cite{van1984properties} exhibiting optical non-reciprocity. Our structures are more amenable to use in small sized integrated systems and unlike YIG, the growth of quantum well devices is established easily through molecular beam epitaxy. We emphasize that leveraging the spin of the electron with a gate field for non-reciprocal photonics remains unexplored heretofore. InSb has been previously explored~\cite{chen2015heat} for its non-reciprocity with emphasis on its gyro-electric behavior, the present work shows that it is possible to design ``multi-gyroic" materials which have non-reciprocity in both the electric and magnetic off-diagonal permeability and susceptibility tensor components. We further note while similar analyses with gyroelectric media exist in literature~\cite{kamenetskii2001nonreciprocal,eroglu2010wave} wherein non-reciprocity has been demonstrated~\cite{lima2011nonreciprocity}, such realizations however are generally incumbent solely upon the external magnetic field and offer no recourse to further modulations via microscopic device rearrangements. Furthermore, throughout the manuscript, we do not invoke the terminology of chirality~\cite{tang2010optical}. Chirality (i.e. traditional optical activity) is a reciprocal phenomenon, and the fields of metamaterials, plasmonics, and chemistry define it as a coupling coefficient of electric and magnetic fields. Gyrotropic  non-reciprocity, in contrast, associated with photon spin inside matter, couples the orthogonal components of the electric (or magnetic) fields. 
\begin{figure*}[ht!]
\center
\includegraphics[width=5.5in]{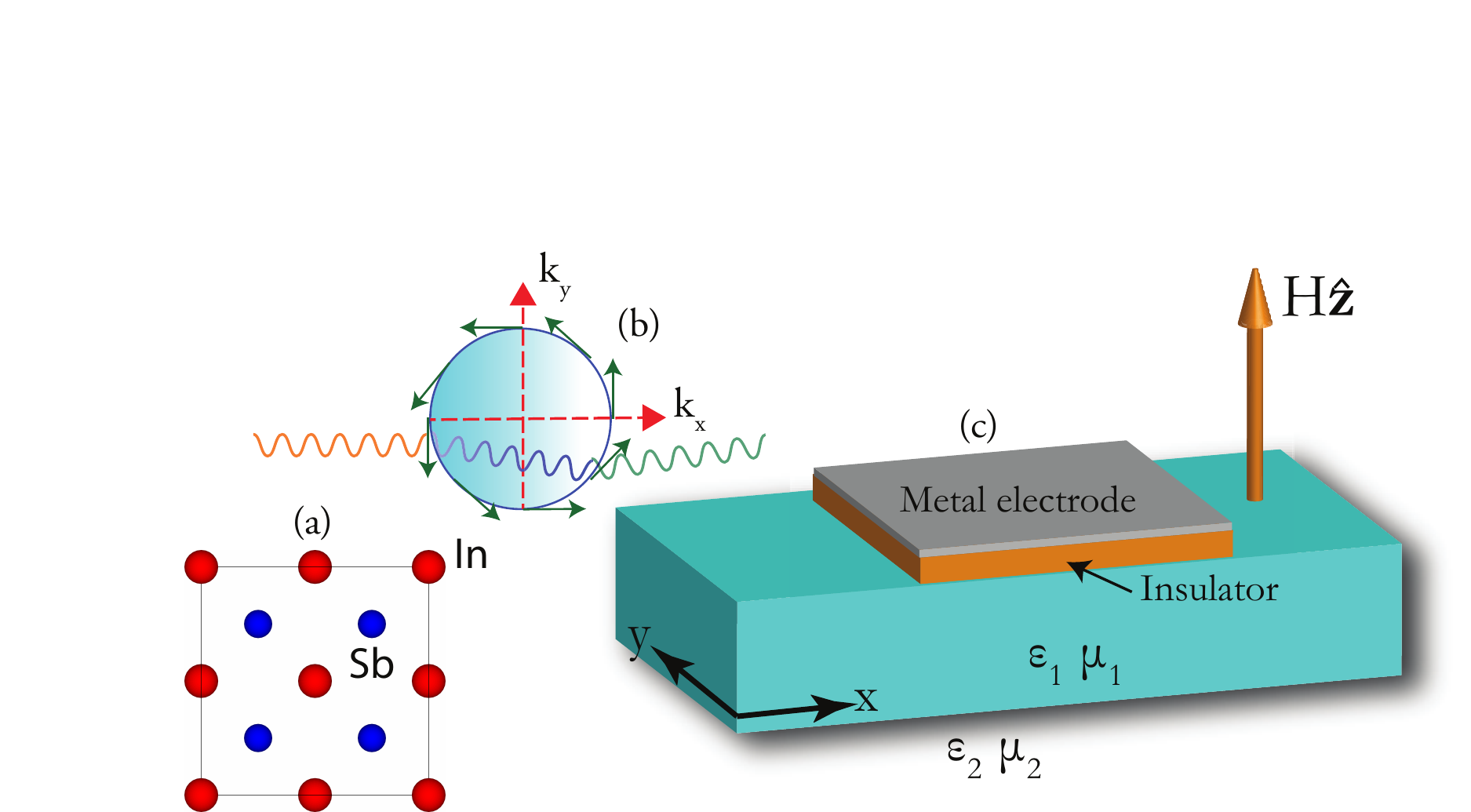}
\caption{The schematic represents the arrangement considered in this work. The left figure (a) shows a unit cell of ferromagnetically doped InSb (red atom denotes In while blue stands for Sb) irradiated with a beam of light (wavy line) that traverses its body and emerges on the opposite side. The passage of the light beam is governed by the constitutive parameters, $ \epsilon_{1}$  and $ \mu_{1} $, of InSb, which is gyrotropic with an inherent magnetization. Note that for gyrotropy to be observed, an out-of-plane magnetic field (Hz) is applied to the device. The permeability tensor in this case is significantly modified by the external Rashba spin-orbit coupling (RSOC) that exists on the InSb slab. The amplitude of transmission of an incident beam through the slab,  marked as an angled wavy blue line in the middle figure (b), is therefore linked to the strength of the RSOC. The RSOC in (b) is identified by its characteristic spin-momentum locking, where the tangential green lines indicate the spin-polarization vectors. The right figure (c) is a possible realization of a gyrotropic and non-reciprocal optical device. It is fitted with a metal gate that allows a dynamic tuning of RSOC, leading to the necessary modulation of the light beam. We elucidate here, via demonstration of such optical control, on an indirect but robust connection between the electron spin and diverse photonic applications.}
\label{sch0}
\end{figure*}

The present work, as mentioned above, combines a large spin-orbit coupling, narrow band gap, and crystalline asymmetry of the target nanostructure, materializing in a significant external Rashba spin-orbit field~\cite{vzutic2004spintronics,manchon2015new}. This effectively changes the material response to an impinging light beam in the presence of an external magnetic field which is discernible from appropriate magneto-optical data. We now give a succinct description of the arrangement on which the theoretical and computational analysis of the latter sections is centred. The model structure is a magnetically-doped InSb (Fig.~\ref{sch0}a) slab with a permanent axis of magnetization $\left(\mathbf{M}\right)$ normal (aligned to the \textit{z}-axis) to the \textit{x-y} plane and forms the optically active component. The slab (Fig.~\ref{sch0}a) is also placed under an external magnetic field parallel to $ \mathbf{M} $ while a gate electrode is affixed to the top. The non-reciprocity of the magnetized InSb slab is captured by the non-zero off-diagonal elements in the permeability $\left(\mu\right)$ tensor matrix. However, beyond the influence of the magnetic fields, the extent to which such non-reciprocity manifests, is also functionally dependent on the gyromagnetic ratio $\left(\gamma = g\left(e/2m_{e}^{*}\right)\right)$. Here, $ e $ stands for the electronic charge and $ m_{e}^{*} $ is the effective electron mass. The \textit{g}-factor, therefore, evidently via $ \gamma $ determines the solutions to Maxwell equations that govern the light-matter interaction in this setup. The middle figure (Fig.~\ref{sch0}b) denotes this process wherein a tailor-able \textit{g}-factor arises as the light beam propagates through a medium with significant external Rashba field identified through the spin-momentum locked states on a equi-energy circular contour. As tangible illustrations of such synergy - albeit indirect - between a photon beam and the Rashba spin orbit coupling (RSOC), we show 1) variations in the characteristic magneto-optical measurements (MO), in particular, the Kerr and Faraday rotation with a varying electric field and 2) the Purcell factor of non-reciprocal photon spin-polarized dipole emission.

Briefly, we note that changes to the Rashba coupling parameter $\left(\lambda_{R}\right)$ through a gate electric field and the dispersion relation (through additional confinement and strain etc.) revises the \textit{g}-factor profile; a higher $ \lambda_{R} $ leading to an enhanced value, and revealed as greater Kerr and Faraday rotations.~\cite{argyres1955theory} We also show electron spin control of photon-spin dependent Purcell factor~\cite{novotny2012principles,khosravi2019spin}. Before we proceed to a complete analysis of the \textit{g}-factor engineered non-reciprocal phenomena, a note about the organization of the paper is in order: In Section~\ref{s2} steps are outlined for the \textit{g}-factor calculation beginning with the model Hamiltonian for the InSb slab; this is followed by a quantitative discussion on electron spin-orbit coupling governed Kerr and Faraday rotations that characterize the viability of non-reciprocity driven magneto-optical devices (Section~\ref{s3}). The Purcell factor, and its numerical determination is taken up next in Section~\ref{s4} and we close by summarizing the key findings in Section~\ref{s5} that also touches upon the possibilities of extending the current work to include aspects of material and structural optimization. 

\vspace{0.25cm}
\section{Theory}
\label{s2}
\vspace{0.25cm}
The basis of all calculations presented in this paper begins with two essential steps : 1) Constructing the permeability $\left(\mu\right)$ tensor matrix that ties its behaviour to the extrinsic Rashba spin-orbit coupling and 2) band dispersion of the two-dimensional (2D) FM. In this section, their analytic expressions are presented in the same order below. Note that at this stage the steps are generalized and no target material is specified; however, we will allude to possibilities during a numerical evaluation of the $ \mu $ matrix and the overall band dispersion later in the manuscript.

We begin by writing the Landau-Lifshitz equation that governs all magnetization $\left(\mathbf{M}\right)$ behaviour in a magnet. In presence of Gilbert damping, and in an external magnetic field $\left(\mathbf{H}\right)$ it takes the form~\cite{lakshmanan2011fascinating,tserkovnyak2002enhanced}
\begin{equation}
\dfrac{\partial M}{\partial t} = \gamma\mu_{0}\left(M \times H\right) + \dfrac{\alpha\gamma}{M}\left(M \times \left(M \times H\right)\right),
\label{lleq}
\end{equation}
where, 
\begin{equation}
\gamma = \dfrac{ge}{2m_{e}^{*}}.
\label{gfaceq}
\end{equation} 
In Eq.~\ref{lleq}, $ g $ is the Lande factor, $ m_{e}^{*} $ is the electron's effective mass, and $ \alpha $ is the Gilbert damping. The magnetic pemeability in vacuum is $ \mu_{0} $. Without loss of generality, we let the magnetic field vector point along the \textit{z}-axis and superimpose a small and identically directed \textit{ac}-field, $ H^{'}\exp\left(i\omega t\right) $. The \textit{ac}-field imparts a frequency dependence to the structure of the $ \mu $ tensor matrix. Analogously, the $ \mathbf{M} $ vector is also assumed to point along the \textit{z}-axis in addition to an induced \textit{ac}-component, $ M^{'}\exp\left(i\omega t\right) $. Inserting the complete expressions for the magnetization and magnetic field in Eq.~\ref{lleq}, the tensor components assume the form~\cite{landau2013electrodynamics}
\begin{equation}
\overline{\overline{\mu}} = \begin{pmatrix} \mu_{xx} & -i\kappa_{xy} & 0\\
i\kappa_{xy} & \mu_{xx} & 0 \\
0 & 0 & \mu_{zz} \\
\end{pmatrix},
\end{equation}
where the individual entries are defined as
\begin{equation}
\begin{aligned}
\mu_{xx}&= 1 + \dfrac{\left(\omega_{0} + i\alpha\omega\right)\omega_{m}}{\left(\omega_{0} + i\alpha\omega\right)^{2} - \omega^{2}},\\
\kappa_{xy}&= -\dfrac{\omega\omega_{m}}{\left(\omega_{0} + i\alpha\omega\right)^{2} - \omega^{2}}.
\label{gyrom}
\end{aligned}
\end{equation}
Finally, $ \mu_{zz} = 1 + M/H $, $ \omega_{m} = \gamma\mu_{0}M $, and $ \omega_{0} = \gamma\mu_{0}H $. This completes the form of the tensor matrix for a gyromagnetic material. A set of remarks is in order here: Firstly, the structure of the $ \mu $ matrix in Eq.~\ref{gyrom}, whose off-diagonal elements vanish (the medium therefore turns isotropic, assuming no gyroelectricity is present) in absence of $ \mathbf{M} $, the intrinsic magnetization vector. Additionally, it is a Hermitian tensor, since $ \mu_{ik} = \mu_{ki}^{*} $. The next comment pertains to the matrix dependence on the electron \textit{g}-factor via the gyromagnetic ratio $\left(\gamma\right)$, a number that is manifestly material-driven; as a case in point, it is determined to be -0.44 for GaAs~\cite{hubner2009temperature} conduction electrons while reaching $ \approx $ 50 in 2D InSb.~\cite{nedniyom2009giant} Notice that the free-electron value of $ g = 2.0023 $ does not apply for a crystal. The \textit{g}-factor of an electron bound to a lattice, \textit{inter alia}, is primarily governed by the intrinsic spin-orbit coupling (soc) and therefore must be computed for each nanosystem including the appropriate quantization effects, which are reflected via the dispersion (electronic) relations through altered (from bulk values) band gaps and effective masses. We will expound on this point in greater detail in the following sub-section and present a path that ties \textit{soc}-effects and their influence on the overall non-reciprocal behaviour.

\vspace{0.25cm}
\subsection{Determination of the g-factor}
\label{s2a}
\begin{figure}
\includegraphics[scale=0.58]{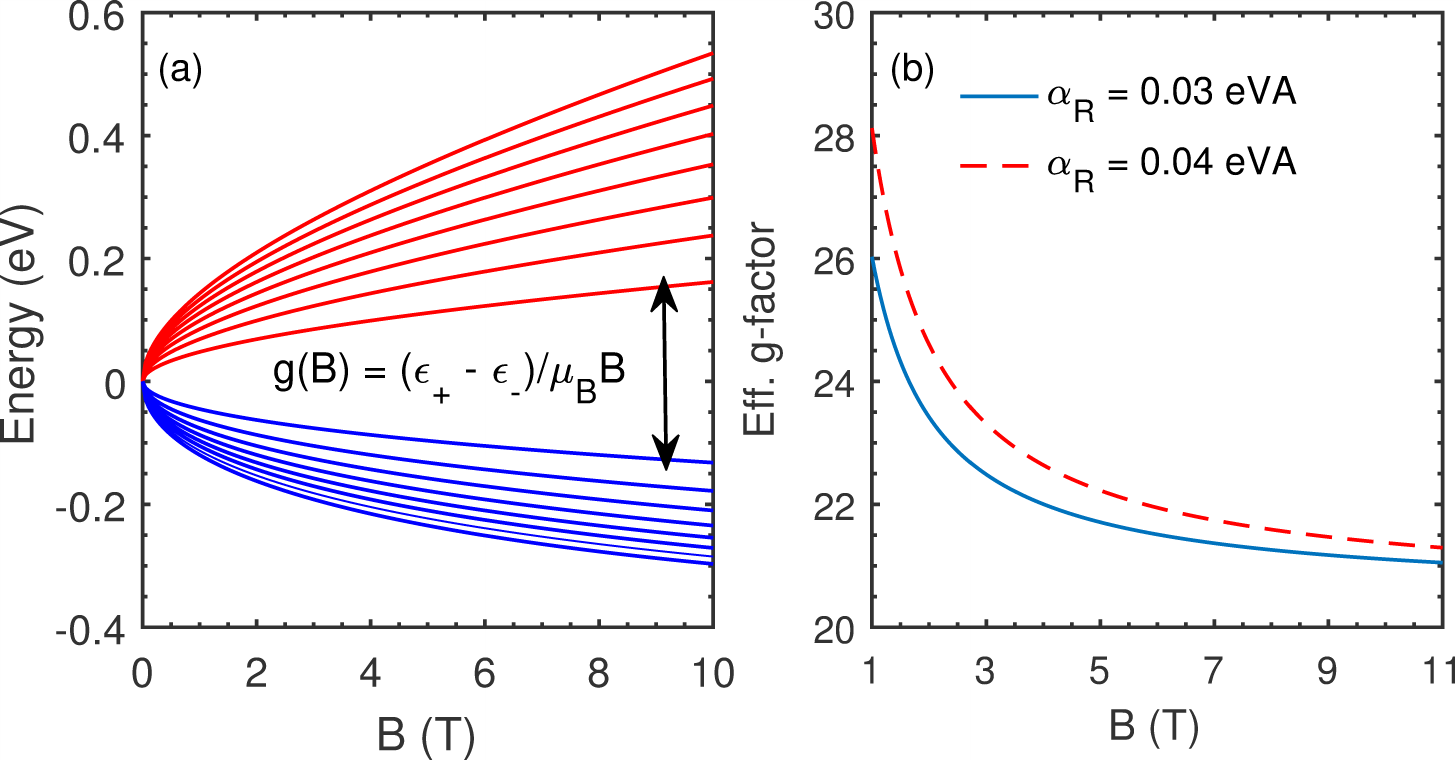}
\caption{The Landau dispersion for the conduction electrons of a $ 15.0\,\mathrm{nm} $ InSb slab for several values of an external \textit{z}-axis directed magnetic field is shown here. The left figure (a) was prepared by diagonalizing the Hamiltonian (Eq.~\ref{hfm3}); the desired InSb band parameters such as the effective mass and the fundamental band gap were obtained from a 8 x 8 \textit{k.p} Hamiltonian adapted for slab-like structures. A note about the band structure calculations and their numerical implementation can be found in the Appendix and Ref.~\onlinecite{sengupta2016numerical}. The upper (lower) set of curves in red (blue) denote the dispersion of the spin-up (down) conduction electrons. The figure on the right (b) is the effective \textit{g}-factor of the conduction electrons computed directly from the Landau dispersion curves. They are shown for two values of the Rashba parameter, a dynamically tunable quantity, an attribute which we harness to describe the coupling between electron spin and optical non-reciprocity in this paper.}
\label{llgeff}
\end{figure}

\vspace{0.25cm}
We remarked above about the functional relationship between the structure of the $ \mu $ tensor and crystal \textit{soc}. In what follows, we make explicit use of band dispersion to formalize this connection. We consider an InSb slab which crystallizes under zinc blende symmetry and displays a substantial RSOC. A minimal Hamiltonian representing the $ \Gamma_{6} $ conduction bands under RSOC is expressed as
\begin{equation}
H_{0} = \dfrac{p^{2}}{2m^{*}} + \lambda_{R}\left(\sigma_{x}k_{y} - \sigma_{y}k_{x}\right),
\label{hfm}
\end{equation}
where $ \lambda_{R} > 0 $ is the Rashba coupling parameter. The effective mass in Eq.~\ref{hfm} is $ m^{*} $. In presence of a \textit{z}-directed magnetic field, carrying out the usual Peierl's transformation, the momentum terms are re-written as : $ \hbar\,k\rightarrow \hbar\,k - e\textbf{A}\left(t\right)$, where $ \mathbf{A} $ is expressed by a Landau gauge of the form $ \left(0,B_{z}x,0\right)$. The momentum terms in Eq.~\ref{hfm}, following this change, can be expressed via creation $\left(a^{\dagger}\right)$ and annihilation $\left(a\right)$ operators, $ k_{+} = k_{x} + ik_{y} = \sqrt{2}/l_{B}a^{\dagger} $ and $ k_{-} = k_{x} - ik_{y} = \sqrt{2}/l_{B}a $, while $ k^{2} $ is now $ 0.5\left(k_{+}k_{-} + k_{-}k_{+}\right) = \dfrac{2}{l_{B}^{2}}\left(a^{\dagger}a + \dfrac{1}{2}\right)$. Here, $ l_{B} = \sqrt{\hbar/eB_{z}} $, the magnetic length. Inserting these transformed momentum representations, the Hamiltonian (Eq.~\ref{hfm}) in matrix form is
\begin{equation}
\begin{pmatrix}
\dfrac{\hbar^{2}}{m^{*}l_{B}^{2}}\left(a^{\dagger}a + \dfrac{1}{2}\right) & i\dfrac{2}{l_{B}}\lambda_{R}a \\
-i\dfrac{2}{l_{B}}\lambda_{R}a & \dfrac{\hbar^{2}}{m^{*}l_{B}^{2}}\left(a^{\dagger}a + \dfrac{1}{2}\right)
\end{pmatrix}.
\label{hfm1}
\end{equation}
The diagonal elements in Eq.~\ref{hfm1} represent a harmonic oscillator. To solve for eigen states, we let the wave function be of the form (assuming translational invariance along the \textit{y}-axis)
\begin{align}
\Psi^{LL}_{n}\left(x,y\right) = \dfrac{\exp\left(ik_{y}y\right)}{\sqrt{\Phi_{n-1}^{2}\left(x\right) + \Phi_{n}^{2}\left(x\right)}}\begin{pmatrix}
\Phi_{n-1}\left(x\right) \\
\Phi_{n}\left(x\right)\\
\end{pmatrix},
\label{wfa1}
\end{align}
where the harmonic oscillator eigen function along the \textit{x}-axis, $ \Phi_{n} = \left(\dfrac{eB}{\pi\hbar}\right)^{1/4}\dfrac{1}{2^{n/2}\sqrt{n!}}\exp\left(-x^{'2}/2\right)H_{n}\left(x^{'}\right) $ and $ x^{'} $ is the short-hand notation for $ \dfrac{\left(x -k_{y}l_{B}^{2}\right)}{l_{B}} $. The Hermite polynomials, $ H_{n}\left(x\right)$, have the usual analytic expression: $ H_{n}\left(x\right) = \left(-1\right)^{n}\exp\left(x^{2}\right)\dfrac{d^{n}}{dx^{n}}\exp\left(-x^{2}\right) $. Employing the standard raising and lowering operator relations, $ a^{\dagger}\Phi_{n} =  \sqrt{n+1}\Phi_{n+1} $ and $ a\Phi_{n} =  \sqrt{n}\Phi_{n-1} $, the Hamiltonian in Eq.~\ref{hfm1} transforms to 
\begin{equation}
H_{0} = \begin{pmatrix}\dfrac{\hbar e B}{m^{*}}\left(n - \dfrac{1}{2}\right) + \Delta & i\lambda_{R}\sqrt{\dfrac{2neB}{\hbar}} \\
-i\lambda_{R}\sqrt{\dfrac{2neB}{\hbar}} & \dfrac{\hbar e B}{m^{*}}\left(n + \dfrac{1}{2}\right) - \Delta
\end{pmatrix}.
\label{hfm3}
\end{equation}
The additional term, $\Delta = \dfrac{1}{2}g_{0}\mu_{B}B $, accounts for the Zeeman-splitting of spin-states in a \textit{z}-axis pointed magnetic field. Note that we set $ g_{0} = 2.0 $ and $ \mu_{B} $ is the standard Bohr magneton. It is now straightforward to diagonalize Eq.~\ref{hfm3} to obtain eigen states for the $ nth $ quantum level; it is simply
\begin{equation}
\mathcal{E}_{n} = \dfrac{\hbar eB}{m^{*}}n \pm \sqrt{\left(\Delta - \dfrac{\hbar eB}{2m^{*}}\right)^{2} + \dfrac{2ne\lambda_{R}^{2}B}{\hbar}}.
\label{eigst}
\end{equation}
The upper (lower) sign is for the spin-up (down) electron. The effective \textit{g}-factor that an electron experiences can then be approximated as 
\begin{equation}
g_{eff} = \dfrac{\mathcal{E}_{1} - \mathcal{E}_{-1}}{2\mu_{B}B}.
\label{geq}
\end{equation}
Notice that we limit our analysis to $ n = 1 $ Landau level for the computation of the effective \textit{g}-factor. In Fig.~\ref{llgeff}, the Landau levels (up to $ n = 8 $) is shown; in addition, the difference in energies between the spin-up and spin-down states for the $ n = 1 $ level is marked on the plot - the precise quantity desired in Eq.~\ref{geq} to ascertain the \textit{g}-factor. 

As a way of elucidation, an additional comment must be included here: The \textit{g}-factor, evidently a function of the Rashba parameter, influences the $ \mu $ tensor (Eq.~\ref{gyrom}) and the concomitant magnetic anisotropy linked optical phenomena. In particular, supplementary degrees-of-freedom in optical manipulation can manifest through alterations made to the strength of the Rashba coupling coefficient, which is $ \lambda_{R} = \lambda_{0}\langle\,E\left(z\right)\rangle $. Here, $ \langle\,E\left(z\right)\rangle $ serves as the average electric field. The material-dependent $ \lambda_{0} $ is given as~\cite{winkler2003spin}
\begin{equation}
\lambda_{0} = \dfrac{\hbar^{2}}{2m^{*}}\dfrac{\Delta_{so}}{E_{g}}\dfrac{2E_{g}+\Delta_{so}}{\left(E_{g} + \Delta_{so}\right)\left(3E_{g} + 2\Delta_{so}\right)}.
\label{rasz}
\end{equation}
In Eq.~\ref{rasz}, the fundamental band gap is $ E_{g} $ and $ \Delta_{so} $ denotes the intrinsic spin-orbit coupling. It is therefore easy to see how a tuning of the essential dispersion parameters - principally, the band gap and electron effective mass - can adjust $ \lambda_{R} $ and thereby the electric and magnetic response of the system. Elucidating further, the electromagnetic response forms the solution to Maxwell's equations that are reliant on the electric permittivity and magnetic permeability of the medium, of which the latter in our case can be transformed via the RSOC-assisted \textit{g}-factor. The set of plots (Fig.~\ref{llgeff}b) reinforces this reasoning. Before we proceed to discuss magneto-optical setups harnessing the embedded utility of the \textit{g}-factor, an explanatory set of statements must be added to dispel any ambiguity: The \textit{g}-factor is typically a tensor quantity and direction-dependent; however, for the case shown here, we assumed the electrons are located at the base of the conduction band which is spherically symmetric $\left(\Gamma_{6}\right)$ allowing a single number to fully represent this inherently tensor quantity. For methods that carry greater rigor and include contributions from higher-energy bands, see for example, Refs.~\onlinecite{hermann1977k,pryor2015atomistic}, a more accurate modeling of the \textit{g}-factor is possible. The Appendix contains a brief note on this point. Lastly, observe that Landau levels derived from a pure parabolic model $\left(\lambda_{R} = 0\right)$ ensures the \textit{g}-factor is independent of the magnetic field - the dependence here otherwise (Fig.~\ref{llgeff}) is simply an outcome of including a linear Rashba spin-orbit Hamiltonian.

\vspace{0.25cm}
\section{Magneto-optical phenomena}
\label{s3}
\begin{figure}
\includegraphics[scale=0.8]{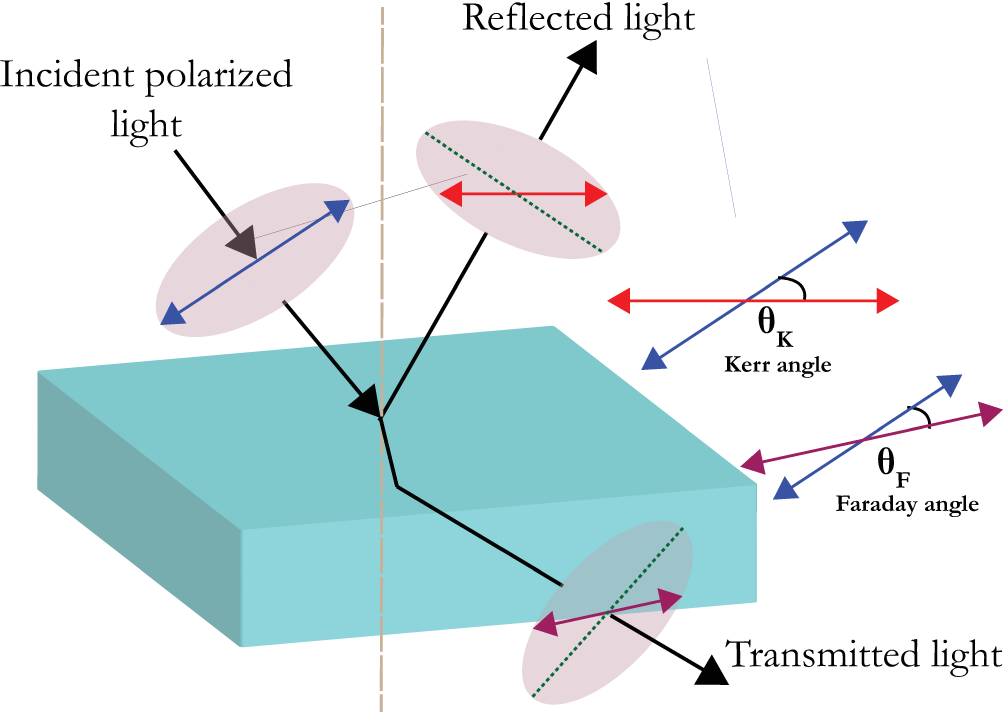}
\caption{The twin optical phenomena of Kerr and Faraday rotation is shown here. The solid lines contained within the ellipses represent the polarization axes which suffer rotation (drawn separately with respective angles marked as $\theta_{K} $ and $ \theta_{F} $) as an incident light beam on the InSb slab is partly reflected and transmitted. Note that this configuration describes the polar magneto-optical Kerr effect (PMOKE) where the magnetization $\left(\mathbf{M}\right)$ is oriented normal to the plane.}
\label{kerrfar}
\end{figure}

\vspace{0.25cm}
A wide variety of functionalities can be accomplished through the inclusion of non-reciprocal photonic devices; however, as we pointed in the opening paragraphs, geometric considerations hinder integration into silicon photonic systems necessitating the need for planar and dimensionally shrunken devices. While magnetic oxide films have been put forward as suitable material systems in this regard, here we seek to explore a class of strongly spin-orbit coupled and narrow band gap zinc-blende materials with embedded magnetic impurities (cf. Fig.~\ref{sch0}). The usefulness of a magneto-optical material is typically gauged  by a figure-of-merit $\left(\xi\right)$ defined as~\cite{jacobs1974faraday} Faraday degree of rotation per dB absorption; more concisely, $ \xi = \theta_{F}/\zeta $, where $ \theta_{F} $ is the Faraday rotation and $ \zeta $ gives the absorption coefficient (per unit length) of the material. It may therefore appear prudent to measure $ \theta_{F} $ and the related Kerr rotation $\left(\theta_{K}\right)$ in the InSb-based setup taken up in this work. The Kerr and Faraday rotation are sketched in Fig.~\ref{kerrfar}. A numerical calculation of $ \theta_{F} $ and $ \theta_{K} $ can be carried out by examining the Fresnel coefficients. In matrix form, for Kerr rotation, we have~\cite{szechenyi2016transfer}
\begin{equation}
\begin{pmatrix}
E_{r}^{p} \\
E_{r}^{s}
\end{pmatrix} = \begin{pmatrix}
r_{pp} & r_{ps} \\
r_{sp} & r_{ss}
\end{pmatrix}\begin{pmatrix}
E_{i}^{p} \\
E_{i}^{s}
\end{pmatrix}.
\label{fresnel}
\end{equation}
Here, $ r_{ss},r_{sp},r_{ps} $, and $ r_{pp} $ are the Fresnel coefficients and the superscript \textit{s}(\textit{p}) stands for \textit{s}(\textit{p})-polarized incident (\textit{i}) and reflected (\textit{r}) electric field. A similar equation can be written connecting the incident and transmitted components of the electric field by introducing another set of Fresnel coefficients, which are, $ t_{ss},t_{sp},t_{ps},t_{pp}$. Note that in this nomenclature, the off-diagonal coefficients $\left(r_{sp},r_{ps},t_{sp},t_{ps}\right)$ point to the inter-mixing of the \textit{s}- and \textit{p}-components. We can numerically ascertain the reflection and transmission behaviour for a completely generalized case of a planar stratified and bianisotropic media that follows the constitutive relations~\cite{ishimaru2003generalized}
\begin{align}
  \mathbf{D} &= \epsb\varepsilon_0 \Ev + \xib\frac{1}{c}\Hv, \nonumber \\
  \mathbf{B} &= \zetab\frac{1}{c}\Ev + \mub\mu_0\Hv
\label{consti}.
\end{align}
For our case, we set the magneto-electric coupling tensors, $\xib $ and $ \zetab $, to zero while $\epsb $ and $ \mub$ are the dimensionless permittivity and permeability tensors. The permeability tensor has non-zero off-diagonal components. The incident, reflected and transmitted fields are then obtained by matching tangential components at the interface, which here straddles the vacuum and the InSb slab. The electric fields must therefore be computed, which we do by first writing the complete wave vector $\left(\kv=(\kv_{\parallel},\pm k_z\right)$ expression for the reflected and incident plane waves consisting of their respective conserved parallel $\left(\kv_{\parallel}\right)$ and perpendicular $\left(\pm k_{z}\right)$ components. The `+' and `-' signs indicate waves propagating away and toward the interface respectively. 

A simple application of Maxwell's equations gives the dispersion relation $k_{\parallel}^2+k_z^2=k_0^2=(\omega/c)^2$, where $k_{\parallel}=|\kv_{\parallel}|$ is real while $ k_{z} $ can assume both real $\left( k_\parallel < k_0\right)$ and complex $\left( k_\parallel > k_0\right)$ values. Note that $ \kv_{\parallel}=(k_{\parallel}\cos\phi,k_{\parallel}\sin\phi) $ where $ \phi $ is the angle subtended by $\kv_\parallel$ with $ x $-axis. With this notation in mind, we substitute the ansatz $ \left[\Ev, \sqrt{\frac{\mu_0}{\varepsilon_0}}\Hv\right]^Te^{i\left(\kv_{\parallel}\cdot\Rv+ k_z z -i\omega t\right)}$ in Maxwell's equations (Eq.~\ref{consti}) to construct the following dimensionless dispersion relation inside the material
\begin{equation}
\text{det}(M+M_k) = 0, \hspace{5pt} \text{for} \hspace{5pt} 
M=\begin{bmatrix} \epsb & \zetab \\ \xib & \mub \end{bmatrix}.
\label{disp1}
\end{equation}
The matrix, $ M_{k} $, is defined by the auxiliary relation
\begin{equation}
\begin{aligned}
M_k&=\begin{bmatrix} 0 & \kb/k_0 \\ 
-\kb/k_0 & 0 
\end{bmatrix}, \\
\kb&=\begin{bmatrix} 0 & -k_z & k_\parallel\sin\phi \\ 
k_z & 0 & -k_\parallel\cos\phi \\ 
-k_\parallel\sin\phi & k_\parallel\cos\phi & 0 
\end{bmatrix}.
\label{disp2}
\end{aligned}
\end{equation}
The $ 6 \times 6 $ material tensor $M$ expresses the constitutive relations and $ M_k $ encapsulates the result of the curl operator on the plane waves. For a completely generalized anisotropic system, we obtain $ k_{z} $ numerically by setting $\text{det}(M+M_k(k_z))=0 $ for a given $(k_\parallel,\phi)$. The fields inside the material are linear combinations of these eigen states described by polarization vectors $ \ev_{j\pm} $ for $ j=\lbrace s,p\rbrace $ given as 
\begin{align}
\ev_{s\pm}=\begin{bmatrix}\sin\phi \\ -\cos\phi \\ 0 \end{bmatrix},
\ev_{p\pm}=\frac{-1}{k_0} \begin{bmatrix}\pm k_z\cos\phi \\ \pm
  k_z\sin\phi \\ -k_\parallel \end{bmatrix}.
\label{spvectors}
\end{align}
The upper (lower) sign is for a wave propagating along the $ +\ev_z \left(-\ev_z\right)$ direction. It is now a straightforward task to calculate the Faraday and Kerr rotation by simply noting the appropriate ratios of the Fresnel coefficients. For Faraday (F) and Kerr (K) rotation, we have~\cite{da2013monolayer}
\begin{equation}
\Theta_{F} = \theta_{F} + i\eta_{F} = \dfrac{t_{ps}}{t_{ss}},\\
\Theta_{K} = \theta_{K} + i\eta_{K} = \dfrac{r_{ps}}{r_{ss}}.
\label{frot}
\end{equation}
where $ \theta_{F/K} $ is the Faraday/Kerr rotation and $ \eta_{F/K} $ stands for the ellipticity of the \textit{p}-polarized wave. Note that the Fresnel coefficients can be in general complex quantities as seen from the form of Eq.~\ref{frot}. Moreover, $ \theta_{F} = \mathfrak{Re}\left[tan^{-1}\left(t_{ps}/t_{ss}\right)\right]$ with a similar relation holding for $ \theta_{K} $, the Kerr rotation.
\begin{figure}
\includegraphics[scale=0.58]{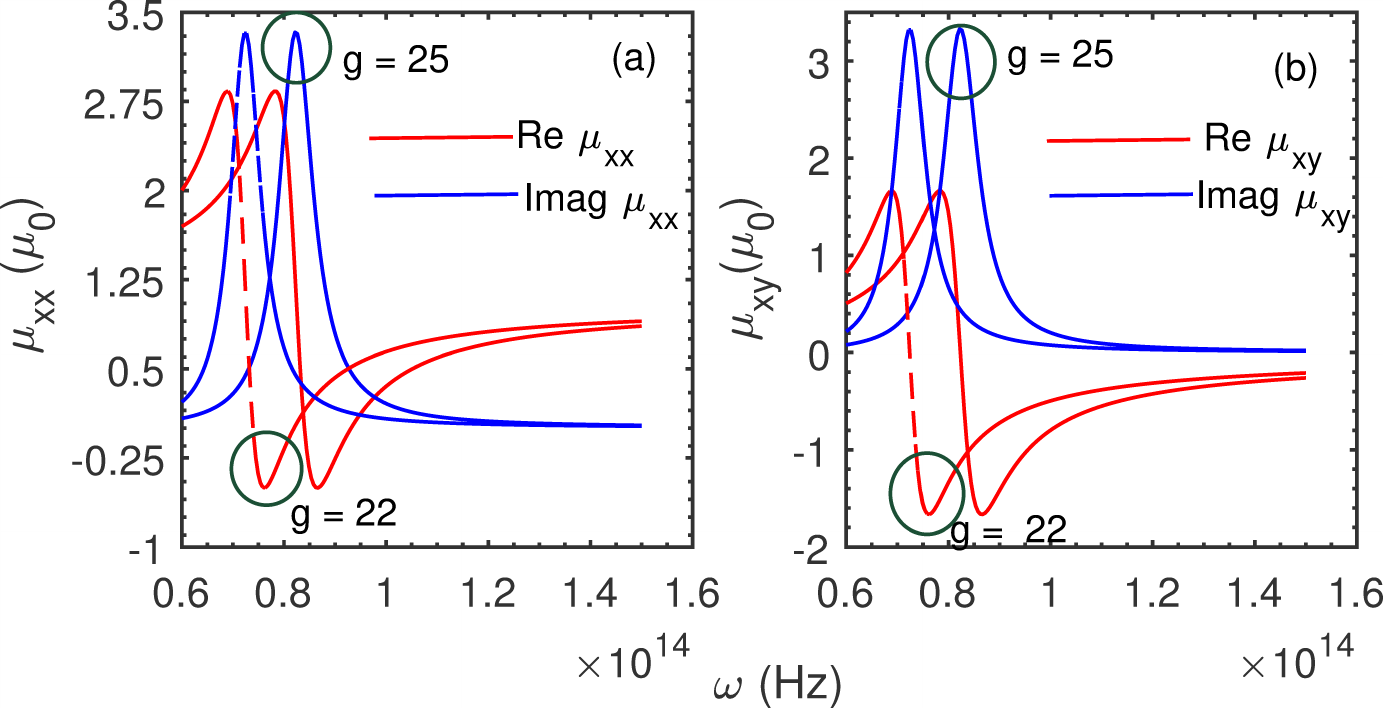}
\caption{The permeability dispersions for two different values of the \textit{g}-factor, where we made use of Eqn.~\ref{gyrom} and set the external \textit{z}-axis directed magnetic field to $ 0.8\, T $ are shown in the above plots. The dispersion curves that use a \textit{g}-factor value of 22 (25) is depicted by a dotted (solid) set of lines. Additionally, the intrinsic magnetization (parallel to the external magnetic field) and the dimensionless Gilbert damping constant were assumed to be $ 0.3\,T $ and $ 0.04\,T$, respectively. The dispersion on the left (a) shows the real and imaginary components of the diagonal elements of the permeability tensor while the right figure (b) furnishes the corresponding curves for the off-diagonal entries. Note that the dispersions for both the diagonal and off-diagonal components besides displaying a functional dependence on the \textit{g}-factor also peak at a resonant frequency. A switch of signs is also observed for a frequency range in both cases.}
\label{mut}
\end{figure}

\begin{figure*}
\includegraphics[scale=0.8]{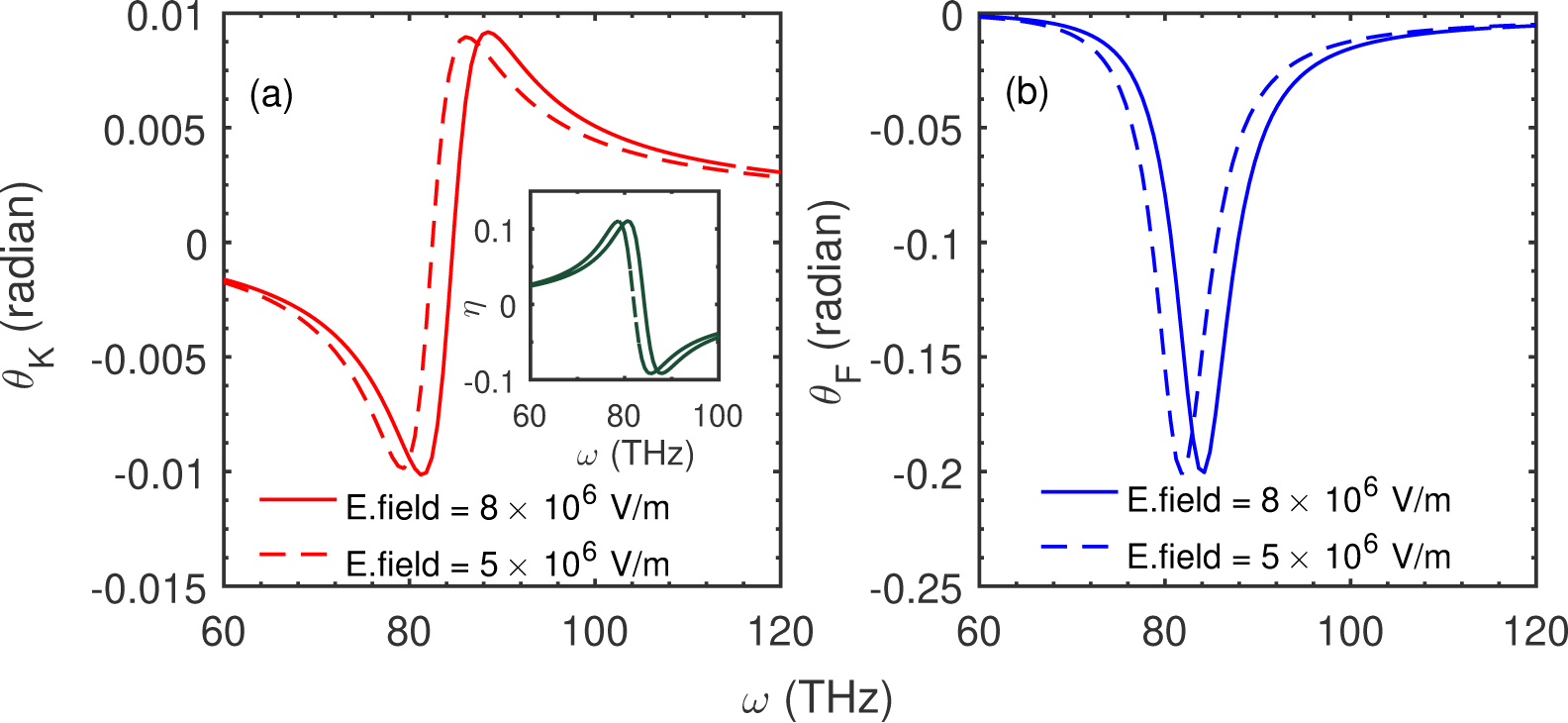}
\caption{We numerically calculate the Kerr (a) and Faraday (b) rotation which arises from reflected and transmitted rays for two gate fields and several incoming frequencies.  The incident light is assumed to make an angle of $ \pi/4 $ with the normal to the plane of incidence. A higher electric field (which augments the g-factor) widens the Kerr rotation angle and also pushes the peak past the one obtained for a lower bias. In addition, the Kerr angle is negative in the same frequency range for which the permeability plots dip below the zero mark (see Fig.~\ref{mut}). The inset in (a) quantitatively assesses the ellipticity of the reflected beam and a profile in agreement with that of the Kerr rotation. The Faraday rotation in (b) which quantifies the plane of rotation of electric field for transmitted waves exhibits a similar behavior for a higher gate bias and records a minimum at the same frequency as noted for its Kerr counterpart. Note that the Kerr and Faraday rotation and the measure of ellipticity are evaluated using the transmission formalism whose governing equations are summarized in Eq.~\ref{frot} in the main text. The material system used in these calculations is a 30.0 nm wide InSb well under an external magnetic field of 0.8 T and intrinsic magnetization of 0.3 T. The Gilbert damping constant, as usual, is set to 0.04.}
\label{compkerr}
\end{figure*}

This brief digression aside, which outlined the steps underpinning a numerical assessment of the Faraday and Kerr rotation, it is now possible to study their dependence on the \textit{g}-factor that impacts the permeability tensor. We show such a calculation in Fig.~\ref{compkerr} and elucidate further: First of all note, that both $ \theta_{K} $ and $ \theta_{F} $ shift with an electric field, an observation easily reconcilable by recalling that the \textit{g}-factor (via the RSOC) undergoes a change leading to a quantitatively different permeability tensor (cf. Fig.~\ref{mut}). It is therefore of interest that an electric (gate) field by acting upon the spin of the electrons for a given magnetic field arrangement (applied and intrinsic) serves as an effective control mechanism to regulate the $ \theta_{F} $-governed figure-of-merit $\left(\zeta\right)$ for magneto-optical devices. It is pertinent to mention here that the key to the adaptability of a non-reciprocal photonic device design is the $ \zeta $ parameter, whose optimization until now has relied on the macroscopic alignment of the total angular momentum of magneto-optical ions (magneto-optic effects are principally an outcome of electronic states with different angular momentum) as a pathway to a high Faraday rotation. A typical arrangement generally brings into play a combined role for the intrinsic spin-orbit coupling of the magneto-optical material and an external magnetic field to achieve a $ \zeta $ commensurate with a level desirable for applications. While in principle, a magnetic field controlled adjustment of material properties is feasible, electromagnetic compatibility and its lack thereof with the adjoining integrated circuitry (in a device environment) makes it a less propitious design guideline. The suggested procedure in this work also involves control of the spin-orbit coupling (external) for a higher Faraday rotation, but with an electric bias that significantly mitigates the severity of electromagnetic incompatibility in case of a magnetic field.

\vspace{0.25cm}
\section{Spin-polarized Purcell effect and the \textit{g}-factor}
\label{s4}
\begin{figure*}[ht!]
\includegraphics[scale=0.7]{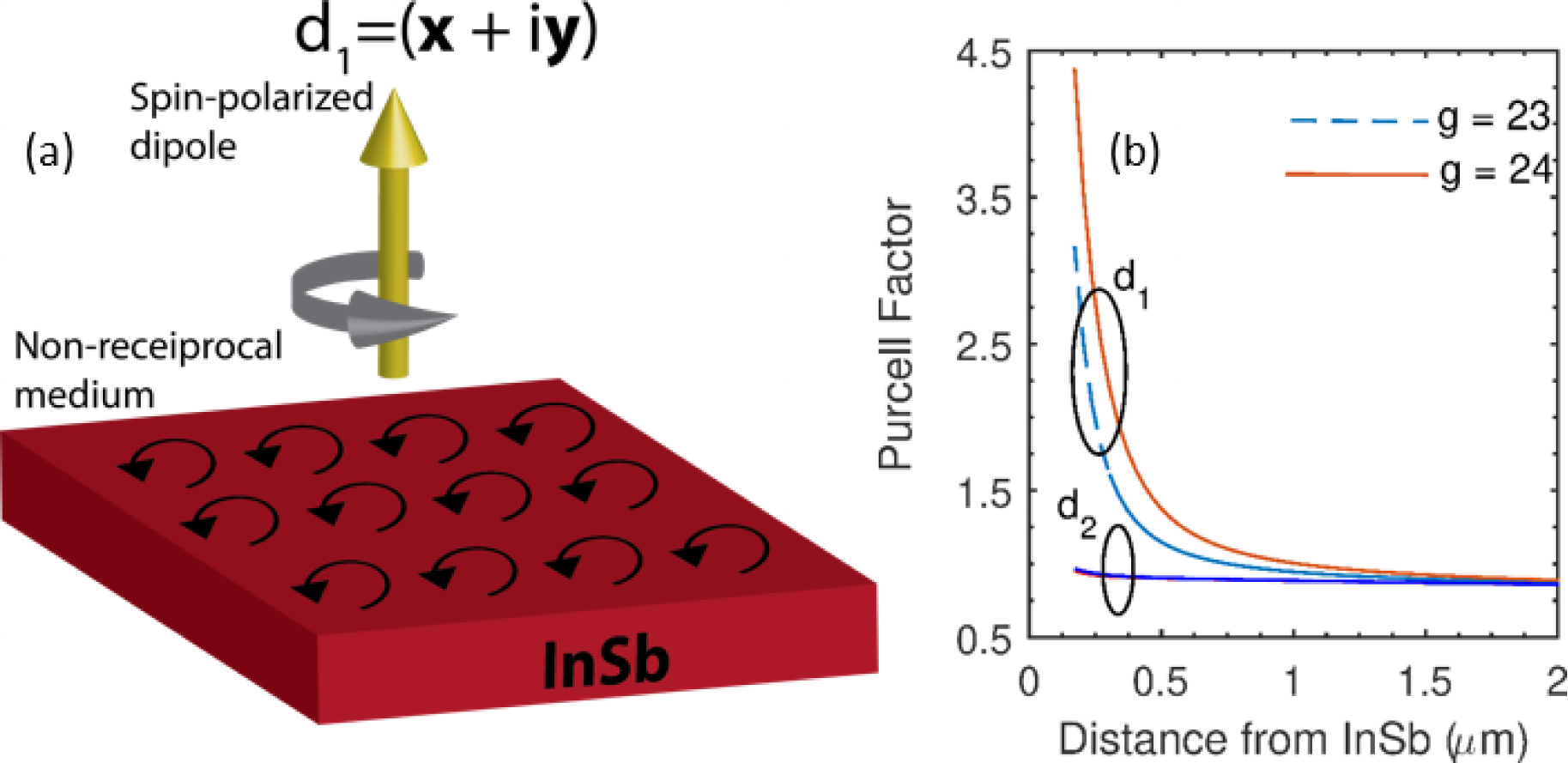}
\caption{The numerically determined Purcell factor (F$_{p}$) for two sets of circularly-polarized dipoles of opposite handedness $\left(d_{1,2} = 1/\sqrt{2}\left[x \pm iy\right]\right) $ placed at a certain distance from a $ 30.0\,\mathrm{nm} $ InSb slab with (assumed) intrinsic magnetization is shown here. We select two values of the \textit{g}-factor for this calculation, where the lower (higher) number corresponds to an electric field of $ 8 \left(5\right) \times 10^{6}\, \mathrm{V/m} $. For the case of $ d_{1} $ dipole, a large enhancement in the Purcell factor is observed, which correlates to an increase in the electromagnetic density-of-states arising from a constructive interference of electric field in the vicinity of the dipole, $ d_{1} $. The placement of the second dipole $\left(d_{2}\right)$ however, in contrast, leads to no significant uptick in the electric field and the F$_{p}$ remains close to unity.  The F$_{p}$, also in case of the $ d_{1} $ dipole reaches a higher value for a g-factor pushed upwards through a stronger gate bias. For smaller distances $\left(z_{0}\right)$ from the slab, a more intense electric field operates that gives rise to a more robust F$_{p}$ uptick; this trend falls off for larger $ z_{0} $ values in agreement with the usual inverse square law for electric fields. Note that the arrow curving around the dipole $ d_{1} $ in  sub-figure (a) represents the emission of a right circularly-polarized light; for $ d_{2} $, the sense of polarization of emitted light is the exact opposite.}
\label{purrfig}
\end{figure*}

\vspace{0.25cm}
We showed how a re-calibration of the permeability tensor via an altered \textit{g}-factor offers promise of tangible dynamic control in magneto-optical measurements. The genesis of such results, which lay in a re-arrangement of the surrounding electromagnetic field, can also be observed in a different setting - the Purcell effect (PE). This effect is characterized by alterations to the spontaneous emission lifetime of a quantum source whose dynamical properties are induced by its interaction with the environment. From an application standpoint, the PE aids in the construction of nano-scale probes and development of newer light sources, for example, lasers and LEDs. The quantitative prediction of PE, therefore, especially where emission-controlled design parameters are of importance. A traditional approach to securing an optimal PE draws upon the geometry and optical attributes of the medium surrounding the emitter, notably, the electromagnetic local density-of-states (LDOS), determined in part, by the constitutive parameters, $ \epsilon $ and $ \mu $. Here, to exemplify the role of the \textit{g}-factor in amendments to the PE, we consider a dipole placed close to the InSb slab and numerically compute the emitter (dipole) decay rate. Nominally, for a dipole moment $ \pv $ located at a distance $ z_{0} $ above the first interface, the PE can be written as~\cite{novotny2012principles} (The frequency and speed of light in vacuum are $\omega $ and $ c $, respectively.) 
\begin{subequations}
\begin{equation}
P = 1 + 6\pi\epsilon_{0}\dfrac{\mathfrak{Im}\pv^*\Gb_{\text{scat}}(z_{0})\pv}{\omega^{3}c^{-3}\vert \pv \vert^{2}}',
\label{peeqn}
\end{equation}
where $ \Gb_{\text{scat}}(z_{0}) $ is the scattered dyadic Green's function of the dipole near the InSb slab that starts at $ z = 0 $ and extends below. We write it as \begin{widetext}
\begin{align}
\Gb_{\text{scat}}(z_{0}) = \frac{i}{2k_z}\int \frac{d^2\kv_{\parallel}}{(2\pi)^2}\bigg[\overbrace{e^{i2k_zz_{0}}[\underbrace{(r_{ss}\ev_{s+}+r_{ps}\ev_{p+})
\ev_{s-}^T}_{\text{reflection of $\ev_{s-}$ wave}} +
\underbrace{(r_{sp}\ev_{s+}+
r_{pp}\ev_{p+})\ev_{p-}^T}_{\text{reflection of $\ev_{p-}$
wave}}]}^{\text{scattered/reflected part $\gb_{\text{ref}}$}}\bigg].
\label{gkz}  
\end{align}
\end{widetext}
\end{subequations}
A plot of the Purcell factor (F$_{p}$) that features the decay rate of the dipole $\left(d_{1} = 1/\sqrt{2}\left[x + iy\right]\right)$ in vicinity of the InSb slab (which serves as a model two-dimensional array of scattering centres) normalized to its value in free space is presented in Fig.~\ref{purrfig}. Clearly, as the \textit{g}-factor is increased, changing the localized electromagnetic setting through the $\mu $ tensor, a stronger field-dipole interaction is revealed as a concomitant rise in the Purcell factor. Further, we carried out the same calculation for a second orientation of the dipole, $\left(d_{2} = 1/\sqrt{2}\left[x - iy\right]\right)$, that yielded no definitive gain for the F$_{p}$. A marginal rise in the decay rate (or equivalently the F$_{p}$) for both values of the \textit{g}-factor points to no significant modification of the localized electric field in presence of the $ d_{2} $ dipole placed above the InSb slab.

We make a comment on the connection of the Purcell effect to the non-reciprocity of the optical medium. Firstly, notice that the scattering matrix in the Purcell formulation identified through the dyadic Green's function (Eq.~\ref{gkz}), say for the dipole $ d_{1} = 1/\sqrt{2}\left[x + iy\right] $, is related to dipole $ d_{2} = d_{1}^{*} $ through the simple relation 
\begin{equation}
\Gb_{\text{scat}}\left(z_{0},d_{1}\right) = \Gb_{\text{scat}}\left(z_{0},d_{2}\right) = \Gb_{\text{scat}}^{T}\left(z_{0},d_{1}\right).
\label{nonrcp}
\end{equation}
The above relation, however, is untrue in a non-reciprocal medium such that the Purcell factors for dipoles $ d_{1} $ and $ d_{2} $ are unequal. Furthermore, since the two dipoles are distinguished through the spins of their emitted light (see Fig.~\ref{purrfig}a and accompanying caption), and display contrasting behaviour, it is conceivable to view this as an instance of photonic spin tied to non-reciprocity.

\vspace{0.25cm}
\section{Final Remarks}
\label{s5}
\vspace{0.25cm} 
We explored the prospects of magneto-optical devices that epitomize the phenomenon of non-reciprocity and showed a newer class of design guidelines can be laid down wherein the electron's spin degree-of-freedom is the primary determinant through the inclusion of the external Rashba spin-orbit-coupling (RSOC) assisted \textit{g}-factor. A set of further advancements can be planned in which the usually weaker Dresselhaus spin-orbit-coupling may actively influence the \textit{g}-factor in tandem~\cite{bulaev2005spin,meier2007measurement} with RSOC, and therefore requires an examination of a large variety of material systems using \textit{ab-initio} techniques. In addition, pursuant to the former objective of suitable candidate materials, a more systematic study of the current setup will aid us to quantitatively correlate (via first-principles simulations) various sample slabs of InSb with strain, magnetized-dopants, defects, and vacancies to magneto-optical phenomena discussed here. Here, we may note that perovskites and its thin film derivatives which are strongly magnetoelectric~\cite{yan2009review,bousquet2011induced} and can carry a robust RSOC is an encouraging alternative to foresee as a starting point for further expanding the design space of magneto-optical structures (and upgrade the FoM $\left(\zeta\right)$ parameter) through a conjoined action of the principles of multi-ferroics and electron spin-orbit coupling. 

The theme of non-reciprocity allied to photon spin was carried over to Purcell factor calculations, where we established using the theory of dyadic Green's function, the decay rate of a dipole held close to an InSb slab. This framework also allows us to assess situations with a randomized configuration of electromagnetic scatters or plasmonic nano-antennas replacing the InSb slab, essentially building a general theory of decay rates in a Purcell factor calculation of emitters (dipoles) near a 2D array of scattering centres. A more comprehensive set of results that suggests structures and emitter orientations maximizing the Purcell effect is planned for a future publication.


\begin{appendices}
\appendix
\renewcommand \thesubsection{\Roman{subsection}}
\titlespacing\section{5pt}{12pt plus 4pt minus 2pt}{0pt plus 2pt minus 2pt}

\vspace{0.25cm}
\section{Band structure calculations}
\begin{figure*}[ht!]
\centering
\includegraphics[scale=0.8]{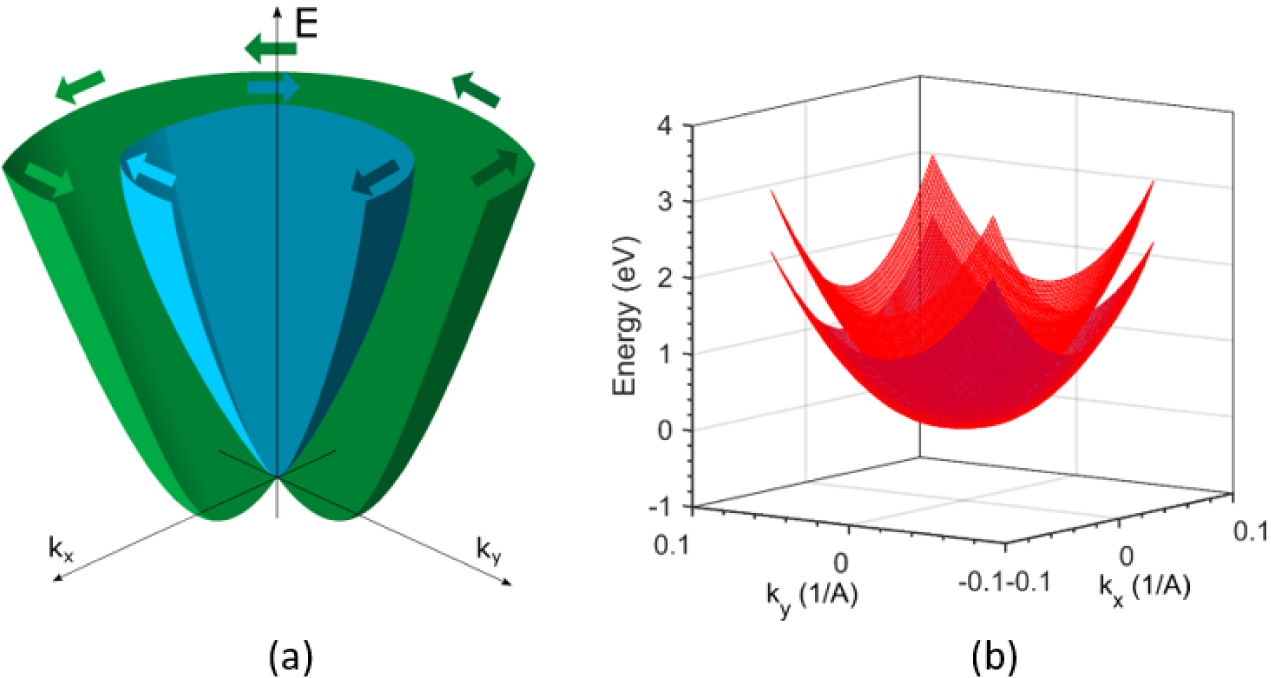}
\caption{The Rashba spin-orbit coupling (RSOC) leads to two non-degenerate Fermi concentric energy contours for the spin-up and spin-down ensemble (a). The right figure (b) shows the band structure of conduction electrons of a $ 6.0\,\mathrm{nm}$ InSb quantum well obtained from a \textit{k.p} calculation. The two “winged-profiles” in the right figure (b) denote the energy contours for the spin-up (higher energy) and spin-down electrons. Notice that InSb is an ideal candidate material to observe RSOC as it satisfies the twin criteria of a large intrinsic spin-orbit-coupling (0.78 eV) and a small band gap (0.43 eV at Brillouin zone centre). In the present case, the Rashba coupling parameter was artificially enhanced to $ 4.0\,\mathrm{eV\AA} $ for a more vivid portrayal of the spin-splitting.}  
\label{schema}
\end{figure*}

\vspace{0.25cm}
We include material that were left out of the main text and brief explanatory notes that clarify and expand on the discussion presented in the paper. The 8-band k.p band structure calculations are performed by discretizing the InSb slab (modeled as a quantum well) on a cubic grid. The quantum well is assumed to be grown along the $ \left[001\right] $-axis. The quantized direction is aligned to $ \left[001\right] $ which is also the \textit{z}-axis. The InSb slab Hamiltonian, 
$ H \left(k_{x}, k_{y}, -i\dfrac{\partial}{\partial z}\right) $, is of size $ 8N_{z} \times 8N_{z} $, where $ N_{z} $ represents the number of discretized points along the \textit{z}-axis. The finite-difference discretization scheme for the 8-band \textit{k.p} Hamiltonian has been explained fully in Ref. 10 of the manuscript. The \textit{k.p} parameters for this work were obtained from I. Vurgaftman \textit{et al.}, Journal of Applied Physics, 89, 5815 (2001). The parameters are also collected in Table~\ref{t1} for easy reference. The conduction band profile of a $ 6.0\,\mathrm{nm} $ InSb quantum well which is spin-split by the Rashba coupling is shown in Fig.~\ref{schema}. In preparing Fig.~\ref{schema}, the effective mass (cf. Eq.~\ref{hfm}) of the conduction electrons were obtained from the eight-band \textit{k.p}-calculation.
\begin{table}[b!]
\caption{8-band k.p parameters for InSb. $E_{v}$, $E_{g}$, $E_{p}$, and V$_{so}$ are in units of eV. The remaining Luttinger parameters are dimensionless constants and the effective mass is in units of the free electron mass.}
\vspace{0.25cm} 
\label{t1}
\begin{tabular}
[c]{|c|c|c|c|c|c|c|c|c|}\hline 
Material & $E_{v}$ & $\gamma_{1}$ & $\gamma_{2}$ & $\gamma_{3}$ & $m^{*}$ &
$E_{g}$ & $E_{p}$ & V$_{so}$\\\hline
InSb & 0.28 & 34.8 & 15.5 & 16.5 & 0.0135 & 0.235 & 18 & 0.81\\\hline
\end{tabular}
\end{table} 
A direct approach to ascertain the \textit{g}-factor (g$_{f}$ in Eq.~\ref{gff}) using \textit{k.p} theory is from the following result
\begin{equation}
g_{f} = g_{0}\left[1 - \dfrac{E_{p}}{3}\left(\dfrac{1}{E_{6c} - E_{8v}} - \dfrac{1}{E_{6c} - E_{7v}}\right)\right].
\label{gff}
\end{equation}
In Eq.~\ref{gff}, $ g_{0} \approx 2 $ is the free electron \textit{g}-factor while the subscripts $ 6c, 7v $, and $ 8v $ designate the symmetries of the bottom (top) of the conduction (valence) bands in a crystal with $ T_{d} $ symmetry. All remote contributions from higher-order bands have been ignored. Note that $ E_{6c} - E_{8v} $ is the fundamental band gap $\left(E_{g}\right)$ and  $ E_{6c} - E_{7v} = E_{g} + \Delta_{so} $. Here, $ \Delta_{so} $ is the splitting from the intrinsic spin-orbit coupling. While in principle, it is possible to derive a similar expression with Rashba coupling term that explicitly accounts for $ E_{g} $, $ \Delta_{so} $, and the effective mass, the approximate estimation procedure outlined in Section~\ref{s2a} indirectly includes the foregoing quantities through the Rashba parameter (cf. Eq.~\ref{rasz}).

Finally, in context of the eight-band \textit{k.p} Hamiltonian based \textit{g}-factor calculations, it is relevant to mention here that the use of only the lowest conduction band is a reasonable approximation for InSb; the next \textit{p}-like conduction band $\left(\Gamma_{7}\right)$ is much above the fundamental direct band gap. A more accurate model, however, must include the $ \Gamma_{7} $ and $ \Gamma_{8} $ conduction bands, for instance, in GaAs, suggesting a 14-band \textit{k.p}-calculation as our starting point. The \textit{g}-factor formula (Eq.~\ref{gff} must reflect this modification through terms of the form.~\cite{hermann1977k}

\end{appendices}

\end{document}